\documentclass[12pt]{iopart}
\usepackage{iopams}  
\usepackage[english]{babel}
\usepackage{graphicx}
\usepackage{cite}
\usepackage{bbm}

\begin{document}

\title{A fast scheme for the implementation of the quantum Rabi model with trapped ions}

\author{ H\'ector M Moya-Cessa}
\address{Instituto Nacional de Astrof\'{i}sica, \'{O}ptica y Electr\'{o}nica \\ Calle Luis Enrique Erro No. 1, Sta. Ma. Tonantzintla, Pue. CP 72840, Mexico }

%
\begin{abstract}
We show how to produce a fast quantum Rabi model with trapped ions. Its importance resides not only in the acceleration of the phenomena that may be achieved with these systems, from quantum gates to the generation of nonclassical states of the vibrational motion of the ion, but also in reducing unwanted effects such as the decay of coherences that may appear in such systems.
\end{abstract}


\maketitle

\section{Introduction}
Trapped ions are considered one of the best candidates to perform
quantum information processing. By interacting them with laser
beams they are, somehow, easy to manipulate, which makes them an
excellent choice for the production of nonclassical states of
their vibrational motion.

The trapping of individual ions also offers many possibilities
in spectroscopy \cite{itano}, in the research of frequency
standards \cite{wineland, bollinger}, in the study of quantum
jumps \cite{powell}, and in the generation of nonclassical
vibrational states of the ion \cite{meekhof}, to name some applications. To make the ions
more stable in the trap, increasing the time of confinement, and
also to avoid undesirable random motions, it is needed that the
ion be in its vibrational ground state which may be accomplished
by means of an adequate use of lasers.

Because of the high nonlinearities of the ion-laser interaction
its theoretical treatment is a nontrivial problem
\cite{vogeliv,vogelii,Wall97,tomb,davidov}. Even in the simplest
cases of interaction one has to employ physically motivated
approximations in order to find a solution. A well-known example
is the {\it Lamb-Dicke} approximation, in which the ion is
considered to be confined within a region much smaller than the
laser wavelength. Other examples are optical and vibrational
rotating wave approximations that are usually performed in order
to find simpler Hamiltonians.

Many treatments also assume a {\it weak coupling} approximation,
such that, by tuning the laser frequency to integer multiples of
the trap frequency results in effective (nonlinear) Hamiltonians
of the Jaynes-Cummings type \cite{cirac2,vogeliii}, in which the
centre-of-mass of the trapped ion plays the role of the field mode
in cavity QED.

Recently it was shown that the quantum Rabi model could be
engineered via the interaction of two laser beams with a trapped
ion \cite{solano}. Pedernales {\it et al.} did it by slightly
detuning both laser beams from the blue and red side bands,
allowing them to construct a Hamiltonian of the Rabi type and
reaching all the possible regimes. However, because the parameters
involved are much smaller than the vibrational frequency of the
ion, $\nu$, the ion can suffer losses that lead to the  decay of
Rabi oscillations \cite{decay,wine3}. There have been attempts to
explain such loss of coherences via laser intensity and phase
fluctuations \cite{Milburn}.

We will show here two approaches in which we can engineer a fast
Quantum  Rabi model (QRM), fast in the sense that the parameters
involved in the interaction may be of the order of $\nu$. Instead
of two off-resonant lasers \cite{solano}, we use only one resonant
beam.

\section{Lamb-Dicke regime}
We can write the Hamiltonian of the trapped ion as
\begin{equation}\label{3010}
     H=H_{\textrm{vib}} + H_{\textrm{at}} + H_{\textrm{int}},
\end{equation}
where  $H_{\textrm{vib}}$ is the ion's center of mass vibrational
energy, $H_{\textrm{at}}$ is the ion internal energy, and
$H_{\textrm{int}}$ is the interaction energy between the ion and
the laser. The vibrational motion can be  approximated by a harmonic
oscillator. Internally, the ion will be modelled by a two level
system. In the interaction between the ion and the laser beam, we will
make the dipolar approximation, so we will write the interaction
energy as $-e\vec{r}\cdot\vec{E}$, where $-e\vec{r}$ is the
dipolar momentum of the ion and $\vec{E}$ is the electric field of
the laser, that will be considered a plane wave. Thus, we write
the Hamiltonian, after an optical rotating wave approximations  as
\begin{equation}\label{3020}
    H= \nu \hat{n} + \frac{ \omega_{0}}{2} \sigma_z + \Omega \left[ e^{i(kx-\omega_l t+\phi_l)} \sigma_+
    + e^{-i(kx-\omega_l t+\phi_l)} \sigma_- \right].
\end{equation}
The first term in the Hamiltonian is the ion vibrational energy;
in the ion vibrational energy, the operator
$\hat{n}=\hat{a}^\dagger \hat{a}$ is the number operator, and the
ladder operators $\hat{a}$ and $\hat{a}^\dagger$ are given by the
expressions
\begin{equation}\label{3030}
    \hat{a}=\sqrt{\frac{\nu}{2}} \hat{x} + i \frac{\hat{p}}{\sqrt{2 \nu}},
\qquad
    \hat{a}^{\dagger}=\sqrt{\frac{\nu}{2}} \hat{x} - i \frac{\hat{p}}{\sqrt{2 \nu}},
\end{equation}
where we have set the ion mass equal to one. Also, for simplicity,
we have displaced the vibrational Hamiltonian by $ \nu/2$, the
vacuum energy, that without loss of generality may be disregarded. \\
The second term in the Hamiltonian corresponds to the ion internal energy;
the matrices $\sigma_z$, $\sigma_+$, and $\sigma_-$ are the Pauli
matrices, and obey the commutation relations
\begin{equation}
   [\sigma_z,\sigma_{\pm}]=\pm 2\sigma_{\pm}, \qquad [\sigma_+,\sigma_-]=\sigma_z,
\label{pauli}
\end{equation}
and $\omega_{0}$ is the transition frequency between
the ground state and the excited state of the ion. \\

By considering the resonant condition,$\omega_0=\omega_l$, and transforming to a picture rotating at $\omega_l$
we obtain the Hamiltonian
\begin{equation}\label{QRM}
    H= \nu \hat{n} + \Omega \left[ e^{i\phi_l} \hat{D}(i\eta)\sigma_+
    + e^{-i\phi_l} \sigma_-\hat{D}^{\dagger}(i\eta) \right],
\end{equation}
where we have defined the so-called Lamb-Dicke parameter
\begin{equation}
    \eta = k \sqrt{\frac{1}{2m\nu}}
\end{equation}
that  is a measure of the amplitude of the oscillations of the ion with respect to the
wavelength of the laser field represented by its wave vector $k$. \\

If we consider the condition $\eta\sqrt{\bar{n}}\ll 1$, where $\bar{n}$ is the average number
of vibrational quanta, we can expand the Glauber displacement operator \cite{Glauber} in Taylor series
\begin{equation}
\hat{D}(i\eta )\approx 1+i\eta \hat{a}^{\dagger }+i\eta\hat{a},
\end{equation}
such that the Hamiltonian (\ref{3020}) reads
\begin{equation}\label{QRM}
    H\approx \nu \hat{n} + \Omega \left[ e^{i\phi_l}\sigma_+
    + e^{-i\phi_l} \sigma_-\right]+i\eta\Omega (\hat{a}^{\dagger }+\hat{a})\left[ e^{i\phi_l} \sigma_+
    -e^{-i\phi_l} \sigma_- \right].
\end{equation}
By setting $\phi_l=\pi$ and making now a rotation around the $Y$ axis (by means of the
transformation $\exp(i\frac{\pi}{4}\sigma_y)$), with $\sigma_y=i\sigma_--\sigma_+$, we
obtain the usual form of  the Rabi Hamiltonian
\begin{equation}
    H=\nu \hat{n}-  \Omega\sigma_z -i\eta\Omega(\hat{a}^{\dagger }+\hat{a})(\sigma_+
    - \sigma_-)
\end{equation}
If we take now $\nu =-2\Omega$, and we use the rotating wave approximation, the Hamiltonian
reduces to the anti-Jaynes-Cummings (AJC) interaction Hamiltonian
\begin{equation}
H=-i \eta  \Omega  \left(\hat{a} \sigma _- -\sigma _+ \hat{a}^{\dagger }\right).
\end{equation}
On the other hand, if we set $\phi_l=0$ and follow the same procedure we obtain
\begin{equation}
H=\nu \hat{n}-  \Omega\sigma_z -i\eta\Omega(\hat{a}^{\dagger }+\hat{a})(\sigma_+
- \sigma_-)
\end{equation}
that, by taking $\nu =2\Omega$, and  using the rotating wave approximation now
reduces to the Jaynes-Cummings (JC) interaction Hamiltonian
\begin{equation}\label{33070}
H=i \eta  \Omega  \left( \hat{a} \sigma _+-\sigma _- \hat{a}^{\dagger }\right).
\end{equation}
Up to here we have been able to construct the Rabi interaction,
equation (\ref{QRM}), with a set of parameters that do not allow
all the regimes because $\eta\ll 1$ only permits the JC and AJC
interactions. However, we should stress that this is a much faster
interaction than the one produced by Pedernales {\it et al.}
\cite{solano} as $\Omega$ is the order of $\nu$.

\bigskip
\section{Fast Rabi Hamiltonian}.

We turn out attention again to the Hamiltonian given in equation
(\ref{QRM}) and set $\phi_l=0$

\begin{equation}\label{3050}
H=\nu \hat{n}+ \Omega \left(
\sigma_{+} \hat{D}(i\eta)+\sigma_{-}
\hat{D}^{\dagger}(i\eta)\right),
\end{equation}
we rewrite equation (\ref{3050}) in a
notation where operators acting on the internal ionic levels are
represented explicitly in terms of their matrix elements, as
\begin{equation}\label{3070}
H=\left(
\begin{array}{cc}
\nu \hat{n} & \Omega \hat{D}\left( i\eta \right)
\\ \Omega \hat{D}^{\dagger }\left( i\eta \right) & \nu
\hat{n}
\end{array}
\right)
\end{equation}
and consider now the unitary operator \cite{dutra,segundo}
\begin{equation}\label{3075}
T=\frac{1}{\sqrt{2}}\left(
\begin{array}{ll}
\hat{D}^{\dag }\left( i\eta/2\right) & \hat{D}\left( i\eta/2\right) \\
-\hat{D}^{\dag }\left( i\eta/2\right) & \hat{D}\left(
i\eta/2\right)
\end{array}
\right).
\end{equation}
 It is possible to check after some
algebra that
\begin{equation}
{\mathcal H}_{\textrm{QRM}}=THT^{\dagger }=\left(
\begin{array}{cc}
\nu \hat{n}+\Omega +\frac{\nu \eta ^{2}}{4} & \frac{\iota \eta \nu }{2}
\left( \hat{a} - \hat{a}^{\dag }\right) +\frac{\delta }{2} \\ \frac{\iota \eta
\nu }{2}\left( \hat{a} - \hat{a}^{\dag }\right) +\frac{\delta }{2} & \nu
\hat{n}-\Omega +\frac{\nu \eta ^{2}}{4}
\end{array}
\right),
\end{equation}
that, after returning to matrix notation, reads
\begin{equation}\label{3080}
{\mathcal H}_{\textrm{QRM}}=\nu \hat{n}+\Omega \sigma
_{z}+\frac{\iota \eta \nu }{2} \left( \sigma _{+}+\sigma
_{-}\right) \left( \hat{a}-\hat{a}^{\dag }\right)  +\frac{\nu
\eta^{2}}{4},
\end{equation}
that is nothing but the  quantum Rabi Hamiltonian plus a constant
term that can be disregarded. A solution for this model has been
given recently by Braak \cite{Braak}

It should be stressed now that in the above Hamiltonian we have not made any
assumptions on the parameters $\Omega$ and $\eta$.

The transformation (\ref{3075}) has already been used to find
(families of) exact solutions to the QRM \cite{moya}. It has been
also used to implement fast quantum gates in trapped ions
\cite{Jonathan}.

This correspondence is very useful, since it enables one to map
interesting properties of each model onto their counterparts in
the other. For instance  ways of realizing substantially
faster logic gates for quantum information processing in a linear
ion chain \cite{Jonathan}.
\bigskip

\subsection{Effective Hamiltonian}Now we show how to produce a fast
dispersive Hamiltonian. Pedernales {\it et al. } \cite{solano}
showed that it is possible to build such a Hamiltonian by using two slightly
of resonant laser beams tuned almost to the blue and red
sidebands. However, as the parameters they used are in general
much smaller than $\nu$, the dispersive interaction constant, may
be very small. Here, we take advantage of the fact that the
Hamiltonian given in (\ref{3080}) has not been approximated and
therefore there are no restriction on the values of their
parameters. By transforming the Hamiltonian (\ref{3080}) with the
unitary operators \cite{klimov}
\begin{equation}
\hat{U}_1=
e^{\epsilon_1(\hat{a}^{\dagger}\hat{\sigma}_+-\hat{a}\hat{\sigma}_-)},
\qquad \hat{U}_2=
e^{\epsilon_2(\hat{a}\hat{\sigma}_+-\hat{a}^{\dagger}\hat{\sigma}_-)};
\end{equation}
with $\epsilon_1,\epsilon_2 \ll 1$,
\begin{equation}
\hat{{\mathcal H}}_{\textrm{eff}}=\hat{U}_2\hat{U}_1
\hat{{\mathcal H}}_{\textrm{QRM}}
\hat{U}_1^{\dagger}\hat{U}_2^{\dagger},
\end{equation}
and setting
\begin{equation}
\epsilon_1=\frac{\eta\nu}{2(\nu+2\Omega)}\qquad
\epsilon_2=\frac{\eta\nu}{2(2\Omega-\nu)},
\end{equation}
remaining up to first order in the expansion $e^{\epsilon
A}Be^{-\epsilon A}=B+\epsilon
[A,B]+\frac{\epsilon^2}{2!}[A,[A,B]]+ ...\approx B+\epsilon
[A,B]$, {\it i.e.} doing a small rotation \cite{klimov}, we
obtain the so-called dispersive Hamiltonian
\begin{equation}\label{effion}
\hat{{\mathcal H}}_{\textrm{eff}}= \nu \hat{a}^{\dagger} \hat{a} +
\Omega \hat{\sigma}_z - \chi_{\textrm{QRM}} \hat{\sigma}_z
(\hat{a}^{\dagger} \hat{a}+\frac{1}{2})
\end{equation}
where the effective interaction constant has the form
\begin{equation}
\chi_{\textrm{QRM}} =\frac{2\eta ^{2}\nu ^{2}\Omega }{4\Omega ^{2}-\nu
^{2}}.\end{equation}

\bigskip
\section{Conclusions} 
Note that most regimes may be achieved with this {\it fast} treatment: Jaynes-Cummings and anti Jaynes-Cummings were produced with the first method ($\eta \ll 1$) and may also be produced with the last one, where the decoupling regime may also be achieved 
$\nu \gg \eta\nu/2 \gg 2\Omega$, the two-fold dispersive regime where $\eta\nu/2 < \nu, 2\Omega,|2\Omega-\nu|, @2\Omega+\nu|$
may be considered, etc. It should again  be stressed that, because decay actually happens in ion-laser interactions \cite{decay,Milburn} it is of great importance to have fast interactions \cite{Jonathan} in order to minimize such unwanted interactions that avoid the generation of nonclassical states, quantum gates, and other important quantum effects.

\end{document}